\documentclass{elsart}

\usepackage{graphicx}
\usepackage[latin1]{inputenc}
\usepackage{amsmath}
\usepackage[dvips]{color}

\begin{document}

\begin{frontmatter}

\title{Yield scaling, size hierarchy and fluctuations of observables in 
fragmentation of excited heavy nuclei.}

\author[ipno]{N.~Le~Neindre},
\author[ipno]{E.~Bonnet},
\author[ganil]{J.P.~Wieleczko},
\author[ipno]{B.~Borderie},
\author[lpc]{F.~Gulminelli},
\author[ipno]{M.F.~Rivet},
\author[lpc]{R.~Bougault},
\author[ganil]{A. Chbihi},
\author[cea]{R.~Dayras},
\author[ganil]{J.D.~Frankland},
\author[ipno,cnam]{E.~Galichet},
\author[ipnl]{D.~Guinet},
\author[ipnl]{P.~Lautesse},
\author[lpc]{O.~Lopez},
\author[gsi,krakow]{J.~Lukasik},
\author[lpc]{D.~Mercier},
\author[ganil,laval]{J.~Moisan},
\author[ganil,buca]{M.~P\^arlog},
\author[napoli]{E.~Rosato},
\author[laval]{R.~Roy},
\author[gsi]{C.~Schwarz},
\author[gsi]{C.~Sfienti},
\author[lpc]{B.~Tamain},
\author[gsi]{W.~Trautmann},
\author[varso]{A. Trzcinski},
\author[gsi]{K.~Turzo},
\author[lpc]{E.~Vient},
\author[napoli]{M.~Vigilante} and
\author[varso]{B.~Zwieglinski}
\begin{center}
(INDRA and ALADIN collaborations)
\end{center}

\address[ipno]{Institut de Physique Nucl\'eaire, CNRS/IN2P3, Universit\'e 
Paris-Sud 11,\\ F-91406 Orsay cedex, France.}
\address[ganil]{GANIL, CEA/DSM-CNRS/IN2P3, B.P.~5027, F-14076 Caen cedex,
France.}
\address[lpc]{LPC, CNRS/IN2P3, ENSICAEN, Universit\'e de Caen, F-14050 Caen 
cedex, France.}
\address[cea]{CEA Saclay, DAPNIA/SPhN, Orme des Merisiers, F-91191 
Gif sur Yvette cedex, France.} 
\address[cnam]{Conservatoire National des Arts et M\'etiers, F-75141 Paris 
cedex 03, France.}
\address[ipnl]{Institut de Physique Nucl\'eaire, CNRS/IN2P3, Universit\'e 
Claude Bernard\\ Lyon 1, F-69622 Villeurbanne cedex, France.}
\address[gsi]{Gesellschaft f\"ur Schwerionenforschung mbH, D-64291 Darmstadt, 
Germany.}
\address[krakow]{Institute of Nuclear Physics IFJ-PAN, PL-31342 Krak{\'o}w, 
Poland.}
\address[laval]{Laboratoire de Physique Nucl\'eaire, D\'epartement de Physique, 
de G\'enie Physique et d'Optique,
Universit\'e Laval, Qu\'ebec, Canada G1K 7P4.}
\address[buca]{National Institute for Physics and Nuclear Engineering,
RO-76900 Bucharest-M\u{a}gurele, Romania.}
\address[napoli]{Dipartimento di Scienze Fisiche e Sezione INFN, Universit\`a
di Napoli\\ "Federico II", I80126 Napoli, Italy.}
\address[varso]{The Andrzej Soltan Institute for Nuclear Studies, PL-00681,
Warsaw, Poland.}

\begin{abstract}
Multifragmentation properties measured with INDRA are studied for single
sources produced in Xe+Sn reactions in the incident energy range 32-50 A
MeV and quasi-projectiles from Au+Au collisions at 80~A~MeV. A comparison
for both types of sources is presented concerning Fisher scaling,
Zipf law, fragment size and fluctuation observables. A Fisher
scaling is observed for all the data. 
The pseudo-critical energies extracted from the Fisher scaling 
are consistent between Xe+Sn central collisions and Au quasi-projectiles.
In the latter case it also corresponds to the energy region at which fluctuations
are maximal. The critical energies deduced from the Zipf analysis are 
higher than those from the Fisher analysis.
\end{abstract}
 

\end{frontmatter}

\section{Introduction\label{intro}}

\indent
From the radial dependence of the nucleon-nucleon interaction, 
containing both repulsive and attractive parts, the nuclear phase
diagram is expected
to present a first order liquid-gas phase transition and a 
second order phase transition at the critical point \cite{siemens}. Nuclear 
collisions offer a large panoply of initial conditions that allow 
to probe the properties of excited nuclear systems and to 
deduce the structure of the phase diagram. It is well established that the 
production of many fragments is the 
dominant phenomenon in reactions over a range of incident energy from 
the Fermi energy up to relativistic energies, both in central and 
semi-peripheral collisions and in reactions induced by 
heavy ions or hadrons on heavy 
targets (see \cite{WCI} for a recent and exhaustive review on experimental 
status). In the chapter "Systematics of fragment observables" of 
this review \cite{WCIb}, it is mentioned that fragment 
production is essentially governed by excitation (dissipated) energy,
dynamical effects being responsible for the observed deviations around the 
general behaviour.
This universal phenomenology does not necessarily imply that the 
occurence of fragmentation has to be associated with a unique mechanism, 
nor that a single trajectory is systematically followed across the phase 
diagram.\\ 

\indent
A crucial piece of information in studies of the fragmentation process is 
cluster size distributions. They inform on intrinsic properties of finite 
excited systems such as scaling laws, size hierarchy or large fluctuations 
\cite{Campi,Ma99,MaYG99,francesca99,francescalat,chomaz1,chomaz2,fisher,chomaz3}. 
These behaviours are actively searched for in multifragmentation data since 
on the basis of theoretical grounds they could signal that nuclear systems 
have undergone a phase transition.\\

\indent
Extraction of a scaling law for the cluster size distribution based on Fisher 
Droplet Model has been reported in a wide variety of collisions from 8 
GeV/c $\pi$ and 10.2 GeV/c proton + Au 
to quasi-projectile (QP) events from peripheral 35~A~MeV Au+Au collisions 
\cite{WCI,elliott1,elliott2,phair02,berkenbusch,scharenberg,michela724}. 
From the fitting procedure of fragment yield distributions with the 
Fisher ansatz, it is possible to extract the energy at which the cluster size
distributions follow a power law, and the associated critical exponents. 
In some works, the
interpretation of the observed scaling has been pushed further, claiming that 
thermodynamical variables, like the temperature and density of the system at 
the critical point can be extracted \cite{elliott1}. On the other hand, 
analyses on QP events have shown that the excitation energy corresponding to
this power law is associated to a peak of abnormal
fluctuation in configurational energy which rather 
signals a system in a coexistence region of the phase diagram 
\cite{michela724}. Since both interpretations were derived in two different 
bombarding energy regimes, it is worthwhile to investigate more deeply the 
scaling properties  of the fragment yields on new sets of data.\\

\indent
Based on Lattice Gas Model calculations, it was recently suggested 
\cite{MaYG99} to examine a specific ordering of the cluster size, the 
so-called Zipf law \cite{Zipf}; in this case,  $<$Z$_{2}$$>$ =$<$Z$_{1}$$>$/2,
$<$Z$_{3}$$>$=$<$Z$_{1}$$>$/3, ...,$<$Z$_{n}$$>$=$<$Z$_{1}$$>$/n where n 
is the rank of the n$^{th}$ cluster in an event having M clusters ordered
by decreasing size Z$_{1}$$>$Z$_{M}$. 
According to the model, such a law would be verified at or close to the 
critical point, see also \cite{watanabe}. Experimental 
investigations \cite{Ma04,Ma05} on the disintegration of quasi-projectiles in 
violent $^{40}$Ar+$^{27}$Al, $^{48}$Ti and $^{58}$Ni
collisions at 47~A~MeV have shown that both critical 
behaviour and Zipf law are observed at the same excitation energy 
E$^{*}$=5-6~A~MeV. Thus, the authors conclude that the Zipf law 
could be a reliable  signature to reveal a critical point. However, 
it has been recently pointed out \cite{campi05} that the Zipf law would be a 
direct consequence of a power law in the yield distribution and would not 
bring more information than the observation of a power law. Since the 
conclusions of \cite{Ma04,Ma05} were strongly supported by the fact that a 
large variety of observables shows a maximal fluctuation at an excitation energy 
where the Zipf law is verified, it is interesting to perform such extensive 
study by comparing different centrality conditions and different size domains 
as in \cite{Ma04,Ma05}.\\ 

\indent
Besides these results, a large body of data has shown numerous signatures 
which are compatible with a coexistence phase of the liquid-gas type:
negative branches of the heat capacity 
\cite{michela2,remi00,nicolasthese,michreliability}; enhancement of 
equal-sized fragment partitions \cite{I31-Bor01,I40-Tab03}; flattening of the 
caloric curve \cite{pocho}; bimodality of the distribution of the largest or 
the asymmetry of the two largest fragments  
\cite{borderiephysg,pic06,Tam05,Bonnet05}. Moreover, recent experimental 
studies on central collisions of symmetric systems at bombarding energies 
around the Fermi energy have shown that data are compatible with a scenario 
in which a compression-expansion cycle leads to a spinodal decomposition of 
the system \cite{I40-Tab03,RIVET98,I29-Fra01,I57-Tab05}.\\ 

\indent
Thus, two types of interpretation could be given from the available nuclear 
multifragmentation data. The first one points to a critical phenomenon and the 
second one supports the picture of a system in the liquid-gas 
coexistence region of the 
phase diagram. This may not be contradictory since various processes
might take place due to different explorations of the phase diagram and 
further studies are needed to investigate such a possibility. This situation 
has motivated the present analysis in terms of Fisher scaling, Zipf law and 
fragment observable fluctuations in measurements performed with the INDRA array. 
The data set used for the present analysis concerns fragmentation data from 
two kinds of centrality: quasi-projectiles produced
in peripheral Au+Au reactions at 80~A~MeV 
(denoted Au QP further on in the text) and 
mono-sources formed in central  
Xe+Sn collisions from 32 to 50~A~MeV. In both cases, the sizes of the 
multifragmenting systems are comparable. Preliminary results using 
Fisher and Zipf techniques have already been reported elsewhere 
\cite{nln02,nln05,I52-Riv05}.\\

\indent
This paper is organized as follows: in Sec. 2, we describe succinctly the set-up of
experiments and we present the methods used to select the events; Sec. 3 
shows the results of the data analysis in the framework of the Fisher 
Droplet Model; in Sec. 4 the cluster size hierarchy is studied according to the 
Zipf law; in Sec. 5, fluctuations of cluster size and configurational energy are 
presented; in Sec. 6 we discuss the ensemble of the collected results. Conclusions are 
drawn in Sec. 7.

\section{  Experimental setup and event selection}

\subsection{  Experimental setup}
\indent
The 4$\pi$ multidetector INDRA is described in detail in 
\cite{POU1,POU2}, and only the main specifications are listed here. INDRA is made 
of 336 detection 
cells arranged in 17 rings; the first one ($2^{o}$-$3^{o}$)
is  an array of phoswich scintillators. Rings 2 to 9
(polar angle from $3^{o}$ to $45^{o}$) consist of
three layers comprising an ionization chamber (IoCh) followed by 
a solid state silicon detector (Si) and a cesium iodide scintillator (CsI(Tl)). 
The medium and backward angular ranges ($45^{o}$ to $176^{o}$) 
are covered with IoCh/CsI ensembles. The device provides a 90\% of $4\pi$ 
geometrical efficiency, a charge identification from H to U, and a mass 
resolution up to beryllium. Data presented here have been obtained with 
INDRA installed at GANIL 
for Xe+Sn reactions between 32 and 50~A~MeV and at GSI for Au+Au at 80~A~MeV. 
For the experiment performed at GSI, a $^{197}$Au beam was impinging a 
2 mg/cm$^{2}$ $^{197}$Au thick target. The INDRA configuration 
used at GSI differs from the one used at GANIL only by the detectors of the 
first ring ($2^{o}$-$3^{o}$). At GSI, the phoswich scintillators were 
replaced with 12 Si-CsI(Tl) telescopes, each consisting of a 300 $\mu$m Si 
detector followed by a CsI(Tl) scintillator of 14 cm length. Further details 
of the experimental and calibration procedures can be found in 
\cite{POU1,POU2,Lukasik,Trzcinski}.

\subsection{ Event selection}

\indent
In Xe+Sn reactions, the set of collisions leading to multifragmenting 
mono-sources has been selected requiring that at least 80\% of the total charge 
and momentum were measured.
On this sample we have performed an event by event shape analysis based on
the 3-dimensional kinetic energy flow tensor, calculated in the centre of mass
frame of the reaction. The tensor is built with fragments with Z$\geq$5 and 
starting from M$_{frag}$$\geq$1. We have defined the
$\theta_{flow}$ angle between the beam axis and the eigenvector associated with
the largest eigenvalue of the diagonalized tensor. A flat cos$\theta_{flow}$
distribution
is observed for $\theta_{flow}$$\geq$60$^{o}$ indicating a strong degree of
equilibration. Thus this sample was retained for the analysis.
More details are given in \cite{I28-Fra01}.

\begin{figure}[ht!]
\begin{center}
\includegraphics*[scale=0.7]{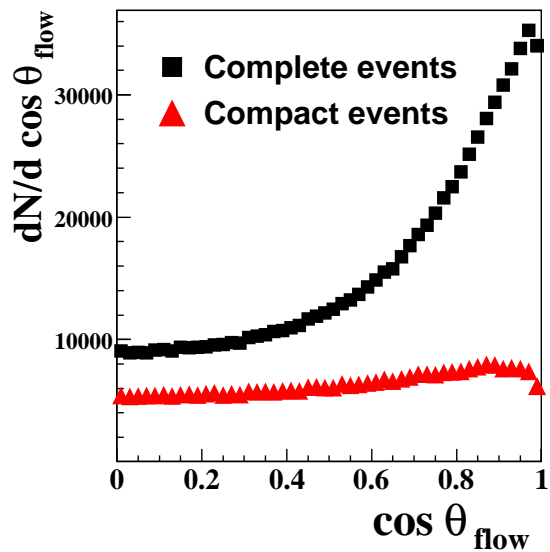}
\caption{\it $\theta_{flow}$ angle distributions for QP formed 
in Au+Au reactions at 80~A~MeV in the QP reference frame. 
Full squares show the flow angle distribution for complete quasi-projectile 
events. Triangles represent the events selected 
according to the procedure explained in the text.}
\label{distrithetaflot}
\end{center}
\end{figure}

\indent
Events comprising at least one fragment with Z $\geq$5 have been considered in 
the analysis of Au QP. We choose this value to be consistent with the Xe+Sn
mono-source analysis.
The kinetic energy tensor is calculated event-by-event and gives the main axis
of the event. A forward hemisphere with respect to this 
main axis can thus be defined. 
The total detected charge Z$_{tot}$ and the 
pseudo total momentum P$_{tot}$ collected in the forward hemisphere are 
calculated by summing up all the charged products having a positive velocity 
component along the main axis. Events satisfying 
0.8$\times$Z$_{proj}$$\leq$Z$_{tot}$$\leq$1.1$\times$Z$_{proj}$ 
and 0.6$\times$P$_{beam}$$\leq$P$_{tot}$$\leq$1.1$\times$P$_{beam}$ 
are kept for the analysis and are called "complete events".
This criterion allows to keep roughly 35\% of the total number of 
measured events.
Then in order to minimize the contribution of dynamical component at 
mid-rapidity \cite{pic06} 
and to well define a quasi-projectile, a compactness criterion
based on velocities is applied.
This method requires that events comprise at least 2 fragments with
Z$\geq$5 and is based on two variables
$\beta_{QP}$ and  $\beta_{rel}$ defined as follows:
$$\beta_{QP}=|\sum \vec{p^{(i)}}.c|/\sum E^{(i)}$$
and
$$\beta_{rel}=\frac{2}{M_{frag}(M_{frag}-1)}\sum_{i > j}|
\vec{\beta^{(i)}}-\vec{\beta^{(j)}}|$$
$\beta^{(i)}$ and $\beta^{(j)}$ are defined in the centre of mass and $E^{(i)}$
represents the total energy of particle $(i)$.

\begin{figure}
\begin{center}
\includegraphics*[scale=0.45]{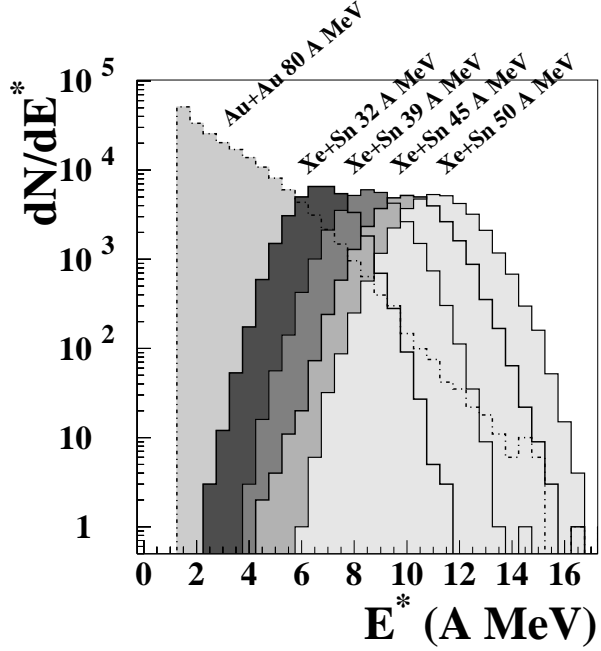}
\caption{\it Excitation energy distributions for Au QP and Xe+Sn
mono-source events between 32 and 50~A~MeV.}
\label{distriexc}
\end{center}

\end{figure}
\begin{figure*}
\begin{center}
\includegraphics*[scale=0.8]{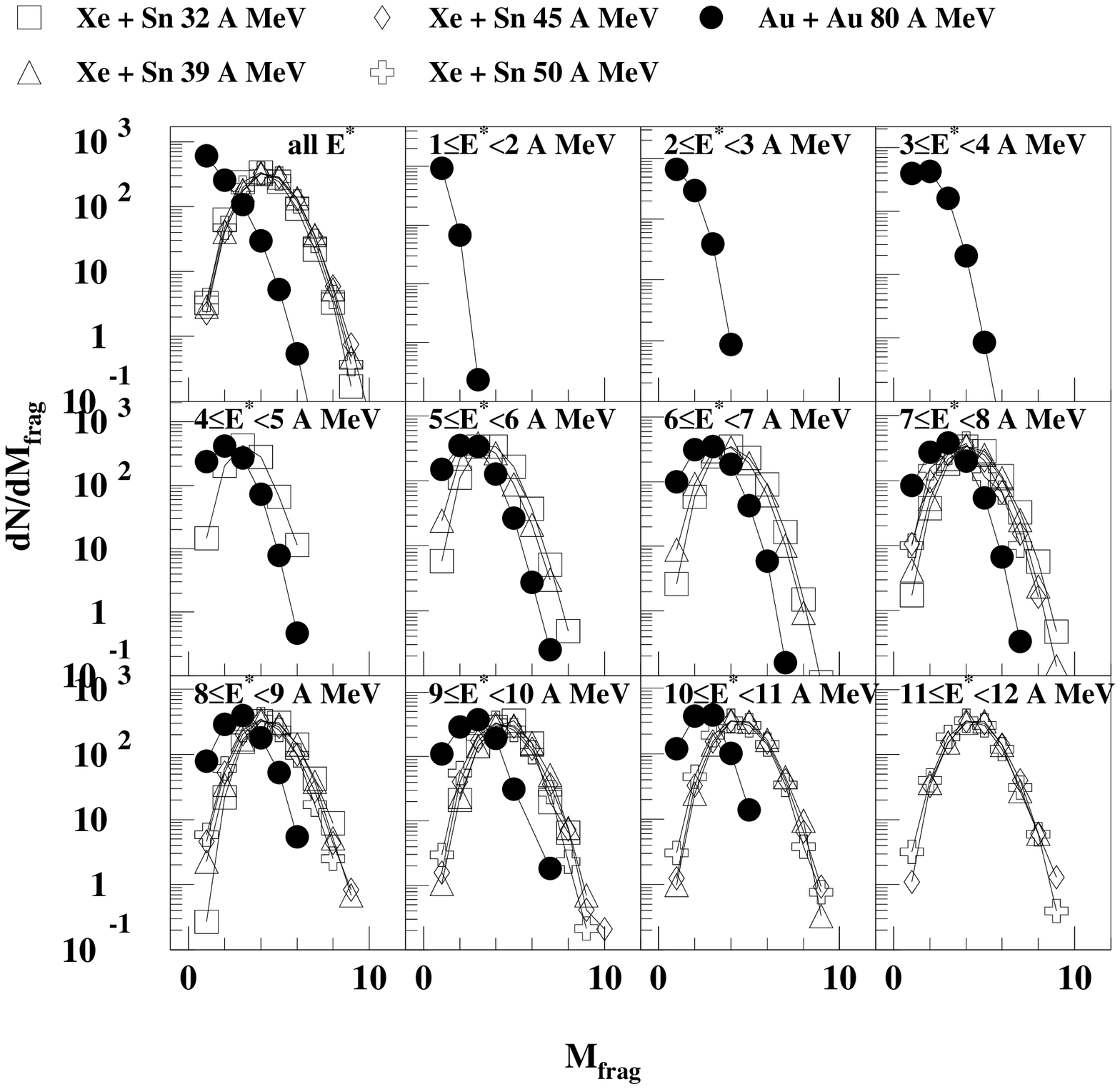}
\caption{\it Top left panel, 
total fragment (Z$\geq$5) multiplicity distributions for Au quasi-projectiles and 
Xe+Sn mono-sources. In the other panels, events are sampled into bins with
1~A~MeV width in excitation energy.}
\label{Multiplicitydist}
\end{center}
\end{figure*}

At this bombarding energy, 80~A~MeV, events are defined as compact if 
$\beta_{QP}$/$\beta_{rel}$ $>$1.5 
\cite{bonnetpaper,bonnetthese}, (24\% of the measured events).
Fission events, characterized by Z$_{1}\times$Z$_{2}$ $\geq$900, 
where Z$_{1}$ and Z$_{2}$ are the first and the second largest fragments,
are rejected \cite{michela1}. 
These events occur mainly at excitation energies lower than 3~A~MeV.
We have checked that the fission fragments have been properly excluded using
a Campi plot as in \cite{mastinu}.
Twice the charge of the light products emitted in the QP forward hemisphere
is added to the sum of the fragment charges to obtain the charge of the 
quasi-projectile Z$_{s}$. Finally we keep events with 
90\%$\times $Z$_{tot}$$\leq$ Z$_{s}$ $\leq$100\%$\times $Z$_{tot}$. 
The considered sample of events exhausts roughly 
5\% of the total measured events:
in the whole procedure, the requirement of having a well 
characterized size is the most constraining.
More details are given in a forthcoming paper \cite{bonnetpaper}.\\

\indent
It is important to check what is the degree of 
equilibration reached in the ensemble of sources under investigation. 
This has been deeply 
studied and widely documented for Xe+Sn mono-sources 
\cite{nicolasthese,RIVET98,I29-Fra01,I57-Tab05,I28-Fra01,I51-Fra05}, therefore 
only the procedure for Au quasi-projectiles is briefly presented here. 
The distribution of the flow angle $\theta_{flow}$, 
recalculated in the source frame, has 
been chosen to illustrate the degree of memory loss of the entrance
channel in the selected collisions. Results of the analysis are presented
in figure \ref{distrithetaflot}. 
Full squares represent the flow angle $\theta_{flow}$ distribution for 
complete quasi-projectile events as defined above. Here one clearly observes 
the typical behaviour of an ensemble of collisions dominated by dynamical 
effects and strongly focused on the beam direction. Triangles are for the 
$\theta_{flow}$ distribution of the selected ensemble.
The $\theta_{flow}$ distribution is flat showing that there is no
longer any privileged 
direction in the disintegration of the system and thus the set of data 
is compatible with the disintegration of an equilibrated source.\\

\indent
In the following, the analysis is performed as a function of the excitation 
energy E$^{*}$ using the calorimetric methods described in 
\cite{nicolasthese,bonnetthese,WCIc}. Accuracy on the calculated 
excitation energy is about 10\% \cite{WCIc}. 
The excitation energy distributions are plotted in 
figures \ref{distriexc} for the events selected in the present analysis. 
For Xe+Sn mono-sources they exhibit a Gaussian shape with a mean value which 
increases with the incident energy. For Au QP, the E$^{*}$ 
distribution extends from 1 to 15~A~MeV.
The shape reflects the range of the impact parameter with the 
highest yield for the peripheral collisions. 
However due to the bombarding energy range (32 to 50 A MeV for Xe+Sn central 
collisions, both types of reaction cover a very similar range in excitation 
energy E$^{*}$.
For each E$^{*}$ it is then possible to calculate some relevant observables such 
as, for example, the size of the source Z$_{s}$ or the fragment multiplicity 
$M_{frag}$. We point out that the mass of the source A$_{s}$ is derived from Z$_{s}$
assuming that its N/Z ratio is the same as that of the complete system.
The average size of the source is around $<$Z$>$$\simeq$80-85 
(with RMS$\simeq$8-9) for the Xe+Sn mono-sources and $<$Z$>$$\simeq$72 
(with RMS$\simeq$5 due to the selection criterion 
90\%$\times $Z$_{tot}$$\leq$ Z$_{s}$ $\leq$100\%$\times $Z$_{tot}$)
for the selected Au QP.\\ 

\indent
Figure \ref{Multiplicitydist} shows the 
fragment multiplicity distributions for 1~A~MeV E$^{*}$ bins.
The top left panel represents the whole distribution integrated over excitation 
energy. The open symbols (full circles) are used for Xe+Sn mono-sources 
(Au QP). In Xe+Sn cases, $M_{frag}$ distributions are weakly
shifted towards higher values when the excitation energy increases
independently of the incident energy.
For excitation energy E$^{*}$ $\geq$4~A~MeV the fragment multiplicity 
distributions in Au+Au are peaked at a slightly smaller value in comparison to 
Xe+Sn and the difference tends to increase as E$^{*}$ increases.
The same trend persists even if $M_{frag}$ is normalized to the size of the 
source. On the top left panel the fragment multiplicity distributions are 
clearly different just because of the different ranges of excitation energy, 
(see figure \ref{distriexc}).\\ 

\begin{figure}[ht!]
\begin{center}
\includegraphics*[scale=0.45]{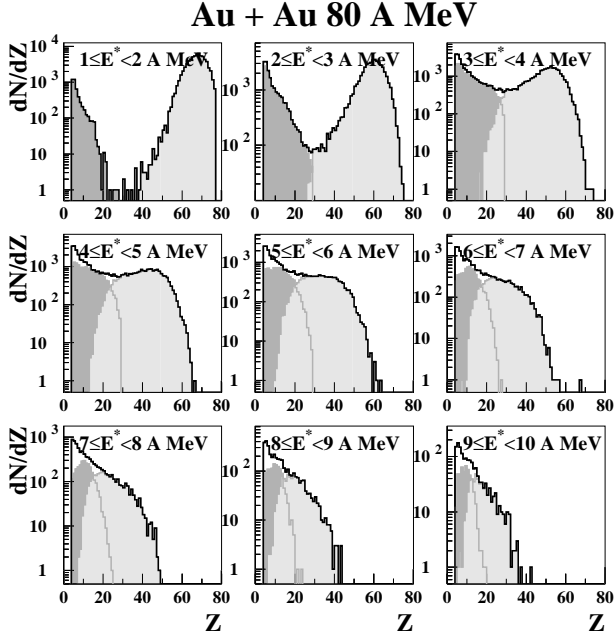}
\caption{\it Charge distributions, starting from Z$\geq$5, for QP formed in Au+Au 
reactions at 80~A~MeV and selected according to the procedure explained in the 
text. Events are sampled into bins with 1~A~MeV width in E$^{*}$
(the bin boundary is indicated). Light (dark) 
grey histograms represent the first (second) largest fragment extracted 
event-by-event. Black histograms are the complete charge distributions.
Removal of fission events is explained in the text.}
\label{distrizau80}
\end{center}
\end{figure}

\begin{figure}[ht!]
\begin{center}
\includegraphics*[scale=0.45]{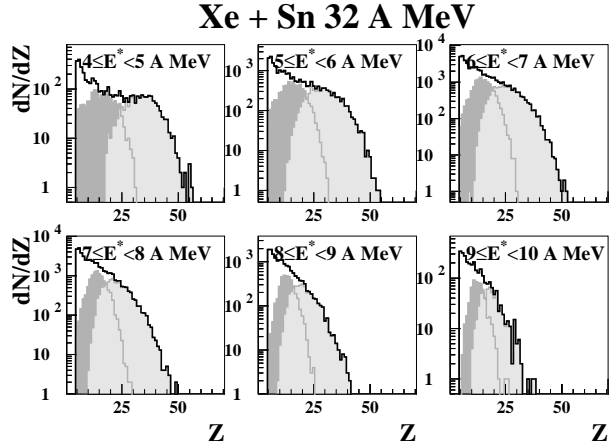}
\caption{\it Charge distributions, starting from Z$\geq$5, for Xe+Sn mono-sources 
at 32~A~MeV. Events are sampled into bins with 1~A~MeV width in excitation 
energy (the bin boundary is indicated). Light (dark) grey 
histograms represent the first (second) largest fragment extracted 
event-by-event. Black histograms are the complete charge distributions.}
\label{distrizxesn32}
\end{center}
\end{figure}

\indent
Figures \ref{distrizau80} and \ref{distrizxesn32} present the excitation energy 
dependence of the charge distributions for quasi-projectiles produced in the 
Au+Au reactions at 80~A~MeV and for Xe + Sn mono-sources at 32~A~MeV, 
respectively.  In both figures light (dark) grey histograms represent the 
first (second) largest fragment extracted event-by-event and black histograms 
are the total charge distributions. In the case of Au QP
at low excitation energy  one observes the  typical U-shape of an 
evaporation process with a big residue and lighter fragments. 
As the excitation energy increases the intermediate charge range is gradually 
populated. At the highest  excitation energy the charge distribution 
falls-off exponentially. The charge distribution in the case of Xe + Sn 
mono-sources presents a similar evolution in the common range of excitation 
energy. Size distributions of the two largest fragments extracted 
event-by-event are also shown. The qualitative behaviour is similar for 
central and peripheral collisions, but from figures \ref{distrizau80} and 
\ref{distrizxesn32} it appears that the shape of distributions of the 
largest fragments are different with a wider distribution in case of the 
QP data. This aspect will be discussed in more details in Section 5.


\section{ Data analysis with the Fisher Droplet Model}
\subsection{Parameterization using Fisher Droplet Model}
\indent
The Fisher Droplet Model (FDM) \cite{fisher,kubo,finocchiaro} describes a gas of noninteracting 
clusters in thermal equilibrium with a liquid fraction. In this approach,
the relative abundance of a cluster containing $A$ nucleons is given by:
\begin{equation}
\eta_{A}=q_{0}A^{-\tau}exp(\frac{A\Delta\mu}{T}-\frac{c_{0}\varepsilon
A^{\sigma}}{T})
\label{fisherequa}
\end{equation}

\noindent
where $\eta_{A}=N_{A}/A_{s}$ is the average number of clusters of mass A per
event, normalized 
to the system size $A_{s}$ and $q_{0}$ is the normalization factor; 
$\tau$ is the topological
critical exponent and $\sigma$ is the critical exponent related to the ratio of
the dimensionality of the surface to that of the volume;
$\varepsilon=(T_{c}-T)/T_{c}$ measures the distance from the critical
temperature; $\Delta\mu$ is the difference in chemical potential from the
liquid phase and $c_{0}$ is the surface energy coefficient. 
Within the model, at the coexistence line $\Delta\mu=0$ and at the critical
point ($\varepsilon=0$) a power law  $q_{0}A^{-\tau}$ is expected for the 
fragment mass distribution.\\

\indent
In the present work we take $T=\sqrt{8 \times E^{*}}$ assuming a Fermi 
gas (the total excitation energy is $E^{*}\times A_{s}$)
as it is done in \cite{michela724}. In the same reference
it has been shown than the values of the critical parameters do
not depend of the details of the caloric curve.
$\Delta\mu$ and $c_{0}$ have been parameterized as polynomials of
order 4 and 1 in E$^{*}$ respectively. The parameters $\tau$,
$\sigma$, $c_{0}$, $\Delta\mu$, $E_{crit}=A_{s}\times T_{c}^{2}/8$ 
and the coefficients of the
polynomial were allowed to vary in order to minimize the $\chi^{2}$. 
This formulation differs slightly from the one used by the authors of
\cite{elliott1,elliott2,michela724}. However, it was checked that $E_{crit}$, 
$\tau$ and $\sigma$ parameters are not very sensitive to the details of the 
scaling function as shown in \cite{michela724}. Indeed this feature makes the 
Fisher scaling analysis a very appealing technique to compare data samples 
and extract universal behaviours.

\subsection{Experimental results}
\begin{figure}[ht!]
\begin{center}
\includegraphics*[scale=0.45]{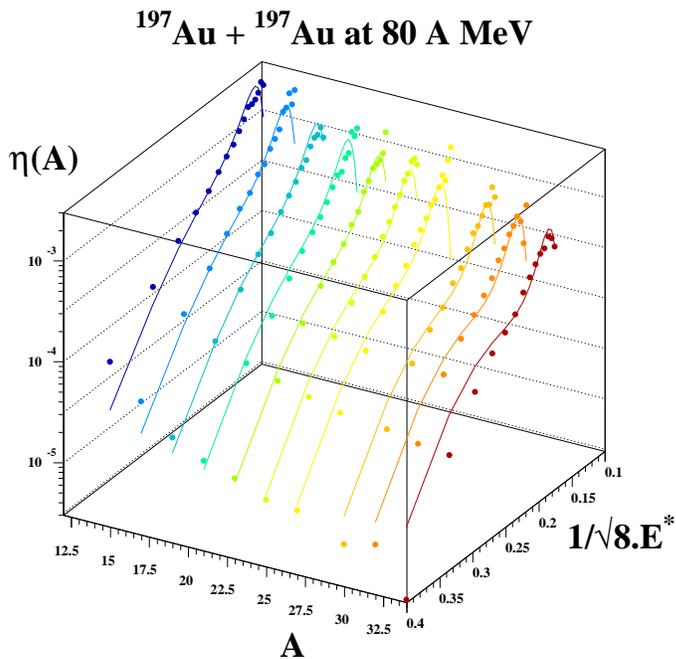}
\caption{\it Yield surface $\eta_{A}$ as a function of A and 
1/$\sqrt{ 8.E^{*}}$. Lines are the results of the fitting procedure to the
Fisher model for charge ranging from Z=6 to 15.}
\label{fitqpau80}
\end{center}
\end{figure}

\begin{table*}
\begin{center}
\caption{\it Values of the critical exponents $\tau$, $\sigma$ and of
E$_{crit}$ extracted from the Fisher Model Analysis for the experimental 
data studied. The value of $\Delta\mu$ and E$_{crit}$ are indicated in 
parentheses when they point out of the range covered by the data set.}
\begin{tabular}{|c|c|c|c|c|c|}
\hline
{\bf System} & $\tau$ & $\sigma$ & E$_{crit} (A MeV)$ & $\Delta\mu$ at E$_{crit}$ & $\chi^{2}$\\
\hline
{\bf Xe + Sn 32~A~MeV} & 2.09$\pm$0.01 & 0.66$\pm$0.01 & 4.50$\pm$0.03 & 0.33 & 2.1\\
\hline
{\bf Xe + Sn 39~A~MeV} & 2.38$\pm$0.02 & 0.66$\pm$0.01 & 4.49$\pm$0.03 & 0.19 & 2.8\\
\hline
{\bf Xe + Sn 45~A~MeV} & 2.40$\pm$0.03 & 0.66$\pm$0.01 & (3.79$\pm$0.03) & (0.90) & 2.7\\
\hline
{\bf Xe + Sn 50~A~MeV} & 2.35$\pm$0.02 & 0.65$\pm$0.01 & (4.23$\pm$0.04) & (-0.02) & 5.7\\
\hline
{\bf Au + Au 80~A~MeV} & 2.56$\pm$0.02 & 0.66$\pm$0.01 & 4.20$\pm$0.03 & 0.24& 9.7\\
\hline
\end{tabular}
\label{tabparaexp}
\end{center}
\end{table*}

\begin{figure}[ht!]
\begin{center}
\includegraphics*[scale=0.45]{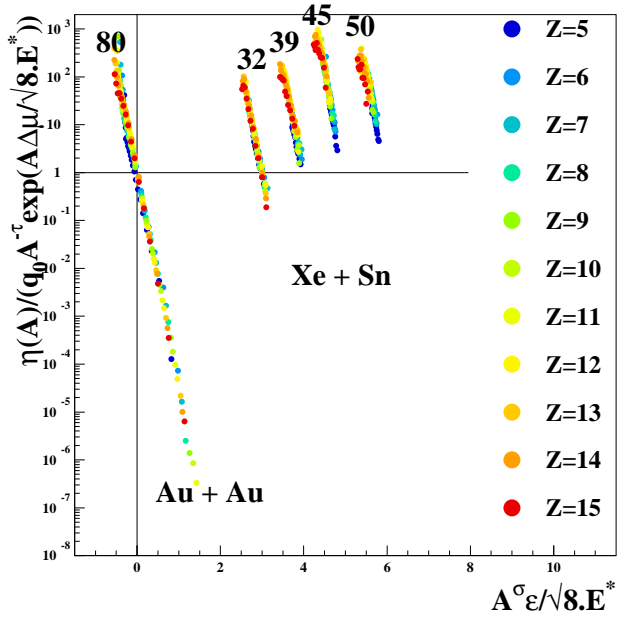}
\caption{\it Scaled yield distributions for Au QP events, 
Xe+Sn mono-sources. Bombarding energies in~A~MeV are 
reported on top of each scaled distribution. A horizontal shift is added for 
each Xe+Sn system (+3, +4, +5, +6 respectively) 
to visualize the scaled distributions.}
\label{scaexp}
\end{center}
\end{figure}

\indent
The results of the fit are shown in figure \ref{fitqpau80} where 
data (circles) and fits (curves) are reported in a three dimensional plot 
($\eta_{A}$, $A$, 1/$\sqrt{8.E^{*}}$). The quality of the fit is 
rather good over a large range of size $A$ and excitation energies E$^{*}$ 
and the main features of 
the excitation functions are reproduced. The next step 
consists in using the scaling properties contained in the Fisher Droplet 
Model parameterization to reduce all the information. This is shown in 
figure \ref{scaexp} where 
$\eta_{A}/(q_{0}A^{-\tau})\times exp(A\Delta\mu/\sqrt{8.E^{*}})$ is plotted against 
$\varepsilon A^{\sigma}/\sqrt{8.E^{*}}$. Now the experimental data from Au QP
collapse in a single line. 
The Fisher Droplet parameterization is applied to the fragmentation
of mono-sources produced in central Xe+Sn collisions.
For each bombarding energy the scaled yields fall on the same line 
(they have been shifted in figure \ref{scaexp} for better visualization) 
meaning that they can be described with the 
exponential dependence of the Fisher Droplet Model.
This scaling feature was also observed in other recent works 
\cite{elliott1,elliott2,michela724,nln02,I52-Riv05,francesca02}. 
The values obtained from the fit, $\tau$, $\sigma$ and E$_{crit}$ are
reported in Table \ref{tabparaexp} together with those of
$\chi^{2}$. The value of $\Delta\mu$ and E$_{crit}$ are indicated in 
parentheses when they are out of the range covered by the data set.
Both $\tau$ and $\sigma$ values are in the range 
predicted by the Fisher Model and are consistent with other experimental 
works. The parameter $E_{crit}$ is around 4-4.5~A~MeV for 
all systems analysed here. This value is commonly obtained in all 
analyses based on
the Fisher scaling technique. In the present data 
$E_{crit}$ are quite similar whatever the centrality of the 
collision and no noticeable role of the entrance channel dynamics on the 
extracted parameters is observed. 
This result agrees with \cite{michela724} in which peripheral and 
central collisions were investigated through the Fisher scaling 
technique.\\ 

\indent
Looking at the results in more details one sees that $\Delta\mu$ is 
slightly positive at $E_{crit}$ (see Tab.\ref{tabparaexp}), which 
means that the size distribution at E$^{*}$=E$_{crit}$ does not follow exactly 
a power law $\eta_{A}=q_{0}A^{-\tau}$. If a 
two parameters ($q_{0}$ and $\tau$)
power law fit is performed on the charge distributions 
at the critical energy, one obtains a $\tau$ value which is much 
lower (1$<$$\tau$$<$1.5) than the one extracted using the scaling analysis 
\cite{nln02}. This difficulty prevents any 
determination of the phase diagram, the location of the coexistence 
region, and the extraction of the critical point,
conversely to what has been done in \cite{elliott1,elliott2}.
We will come back on the interpretation 
of the Fisher scaling in the general discussion of section 6.
   
\section{Cluster size hierarchy and Zipf law}
\indent
Originally the Zipf law was used to analyze the relative population of words
in texts. The frequency of the word is inversely proportional to its rank
in a frequency list \cite{Zipf}. The integer rank n is defined starting from 1 
for the most probable. Later, many applications
of this relationship were made in a broad variety of areas, such as 
city-population 
distributions, sand-pile avalanches, the distribution in strengths of
earthquakes, the genetic sequence, etc. It has been suggested that the 
existence of  similar linear hierarchy distributions in these very different 
fields indicates that the Zipf law is a fingerprint of criticality.
In particular,
recent investigations  with Lattice Gas Model \cite{Ma99} have shown that 
the cluster distribution follows a Zipf law at the critical point. 
In this case the analysis was shifted from the frequency to the cluster size.
This has raised a strong interest around this observable
and a Zipf law has been observed in the  
multifragmentation of a quasi-projecile of Ar in conjunction with other 
signatures of critical behaviour, leading the authors to conclude that the 
nuclear system has been observed at or close to the critical point
\cite{Ma05}.\\

\indent
In the present study we explore the applicability of the Zipf law to heavy 
excited systems of similar size formed either in central or peripheral 
collisions. For each E$^{*}$ bin of 1~A~MeV width, the value of $<$Z$_{n}$$>$ 
is calculated for each rank n. We first consider the fragments with 
Z$\geq$3.
The energy dependence of all $<$Z$_{n}$$>$ is fitted using the formula 
$<$Z$_{n}$$>$ $\simeq$ n$^{-\lambda}$ for n$\leq$6 to have a good 
statistic. This procedure provides the evolution of 
$\lambda$ with E$^{*}$ and allows to localize the excitation energy E$_{Zipf}$
where the exact Zipf law is satisfied ($\lambda$=1).\\

\begin{figure}[!hbt]
\begin{center}
\includegraphics*[scale=0.45]{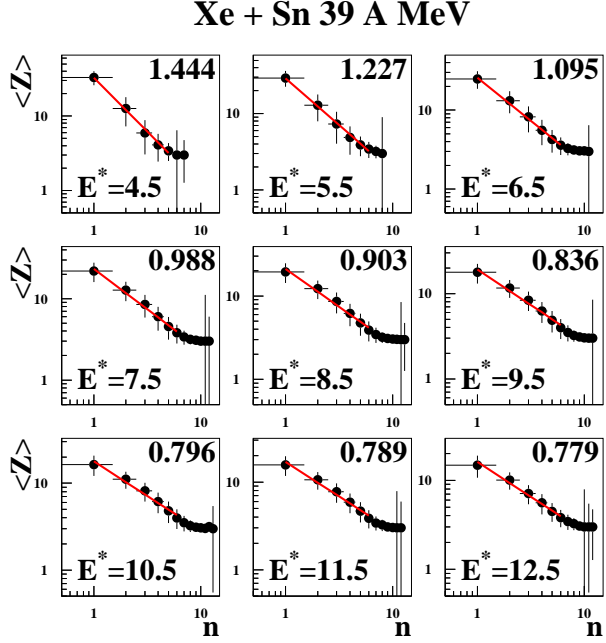}
\caption{\it For Xe+Sn at 39~A~MeV, $<$Z$>$ versus
rank n, in decreasing charge order, for different bins of E$^{*}$
(the mean value of the bin is indicated).
The fit $<$Z$_{n}$ $>$ $\simeq$ n$^{-\lambda}$ is represented by the line and 
the $\lambda$ value is indicated in the right upper part of each panel.}
\label{fitxesn39} 
\end{center}
\end{figure}

\begin{figure}[!hbt]
\begin{center}
\includegraphics*[scale=0.45]{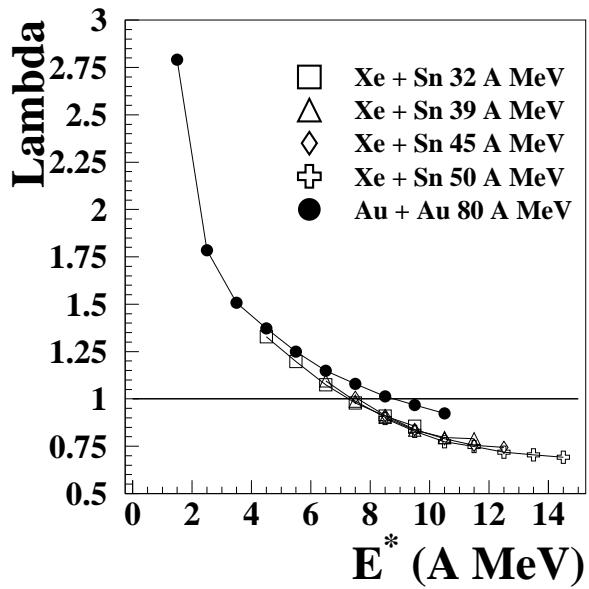}
\caption{\it E$^{*}$ dependence of the $\lambda$ parameter extracted 
from the fit $<$Z$_{n}$ $>$ $\simeq$ n$^{-\lambda}$ for Xe+Sn at  32, 
39, 45  and 50~A~MeV mono-sources and Au QP events at 
80~A~MeV. The Zipf law is satisfied when $\lambda$=1 (horizontal line).}
\label{xesn} 
\end{center}
\end{figure}

\indent
Results are shown on figure \ref{fitxesn39} for Xe+Sn mono-sources at 39 
A MeV. Similar results 
are obtained at other incident beam energies discussed in this article. 
The $\lambda$ values extracted from the fit are reported in the right upper 
part of each panel. Figure \ref{xesn} shows 
the $\lambda$ dependence as a function of E$^{*}$.
The main feature extracted from figure \ref{xesn} is a monotonic decrease of the 
$\lambda$ parameter with increasing E$^{*}$. 
At 32 and 39~A~MeV, the Zipf law is satisfied at 
E$_{Zipf}$=E$^{*}$$\simeq$7.5~A~MeV. As the excitation
energy increases $\lambda$
stays below unity. The same procedure is applied to Au QP. 
The obtained values do not superimpose on the ones extracted from Xe+Sn mono-sources 
and $\lambda$=1 is reached at a higher energy E$_{Zipf}$$\simeq$8.5~A~MeV.\\

\indent
We have also considered the influence of the minimal value of the charge 
used for the analysis. This is all the  more important that the charge 
asymmetry is different in both systems, see next section. 
With increasing minimal charge both curves of figure \ref{xesn} shift
towards lower excitation energy. The shift is even larger for Au QP. Indeed for 
Z$\geq$4 both curves superimpose and we obtain 
E$_{Zipf}$=E$^{*}$$\simeq$6.5~A~MeV.\\

\indent
According to \cite{Ma99} and based on the prediction of the Lattice Gas Model, 
the Zipf energy, E$_{Zipf}$, would be one of the characteristics of the critical 
point. In our data the "critical energies" found, E$_{Zipf}$ and E$_{crit}$,  
using the Zipf-like and Fisher-like analysis are not consistent. 

\section{ Cluster size and kinetic energy fluctuations}
\subsection{Largest and second largest fragment.}

Maximal fluctuations in fragmentation data are actively 
searched for since it is believed that they would be observed 
in the coexistence region or close to the 
critical point \cite{FrancescaAnal}. In this part we will show new results on
properties of the largest fragment, since various experimental and theoretical 
works suggest that the largest fragment can be associated with an order 
parameter for the phase transition \cite{francesca06}.\\

\begin{figure}[!hbt]
\begin{center}
\includegraphics*[scale=0.45]{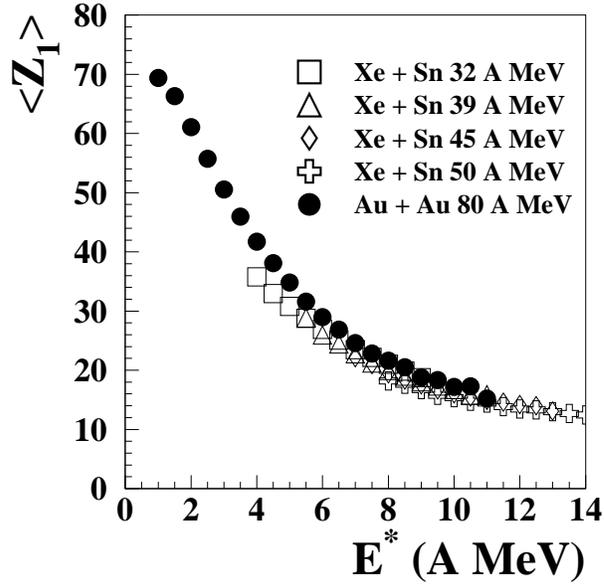}
\caption{\it Evolution of the average charge of the largest fragment 
as a function of the excitation energy E$^{*}$ for Xe+Sn at  
32, 39, 45 and 50~A~MeV mono-sources and Au QP events.}
\label{zbigEstar} 
\end{center}
\end{figure}

\begin{figure}[!hbt]
\begin{center}
\includegraphics*[scale=0.45]{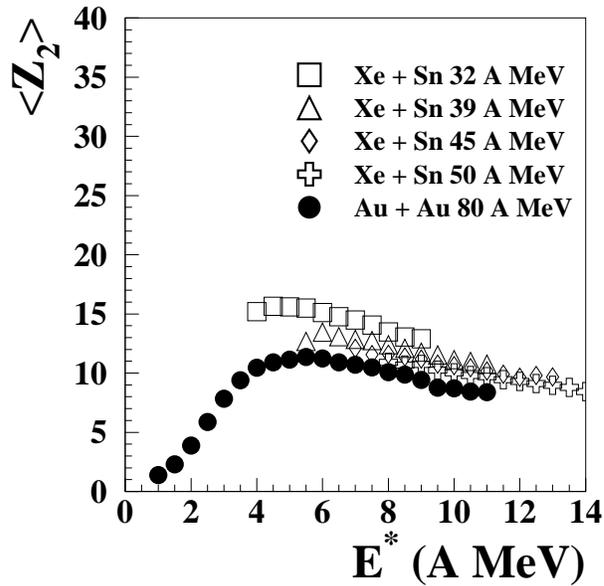}
\caption{\it Evolution of the average charge of the second largest fragment 
as a function of the excitation energy E$^{*}$ for Xe+Sn at  32, 39, 
45 and 50~A~MeV mono-sources and Au QP events.}
\label{zbig2Estar} 
\end{center}
\end{figure}

\indent
Figures \ref{zbigEstar} and \ref{zbig2Estar} show the evolution of the 
average charge of the largest, $<$Z$_{1}$$>$ and second largest, $<$Z$_{2}$$>$,
fragment as a function of the excitation energy E$^{*}$ for 
Au QP and Xe + Sn mono-sources at 32, 39, 45 and 50~A~MeV. 
The source size for Au QP events is $<$Z$_{s}$$>$=72 and for Xe+Sn mono-sources
it has values of $<$Z$_{s}$$>$=83, 81, 79, 79 for 32, 39, 45 and 50
A MeV bombarding energy, respectively. For Au QP, $<$Z$_{1}$$>$ 
decreases from a value very close to the projectile charge for 
E$^{*}$$\simeq$1~A~MeV down to $<$Z$_{1}$$>$=15 at E$^{*}$=12~A~MeV.
Data from Xe+Sn mono-sources are presented as open symbols in figure \ref{zbigEstar}.
Although the Au source is smaller, one observes that 
the corresponding Z$_{1}$ values are slightly larger than the Xe+Sn ones.  
At E$^{*}$=4~A~MeV the difference is around 6 charge units and becomes 
negligible around 10~A~MeV. Thus there is no scaling of the charge of
the largest fragment with the source size.\\

\indent
Figure \ref{zbig2Estar} shows the evolution of the average charge of the second 
largest fragment (including light charged particles),
$<$Z$_{2}$$>$, as a function of excitation energy E$^{*}$ for 
Au QP (filled circles) and  Xe + Sn mono-sources
at 32, 39, 45 and 50~A~MeV (open symbols). For the QP data, one observes a rise 
and fall with a maximum at E$^{*}$$\simeq$5~A~MeV. 
This feature is not surprising;   
it is known that at very low excitation energy, 
in the evaporation regime, $<$Z$_{2}$$>$ is close to one-two. 
Increasing with the opening of the fragment-evaporation channel, $<$Z$_{2}$$>$
reaches a maximum when multifragmentation becomes dominant. It then slowly
decreases similarly to Z$_{1}$ at large E$^{*}$. Thus a maximum is expected for
this observable on a range of E$^{*}$ as it is the case for Au
QP. 
Comparing both data sets at the same E$^{*}$, one observes higher  
$<$Z$_{2}$$>$ values for mono-sources, particularly at 32~A~MeV. 
For higher bombarding energy up to 50~A~MeV, 
all values of $<$Z$_{2}$$>$ roughly collapse
on a single curve and decrease linearly as E$^{*}$ increases. 
Finally, at a given excitation energy the asymmetry between the two largest 
fragments is higher for the Au QP.\\

\begin{figure}[!hbt]
\begin{center}
\includegraphics*[scale=0.45]{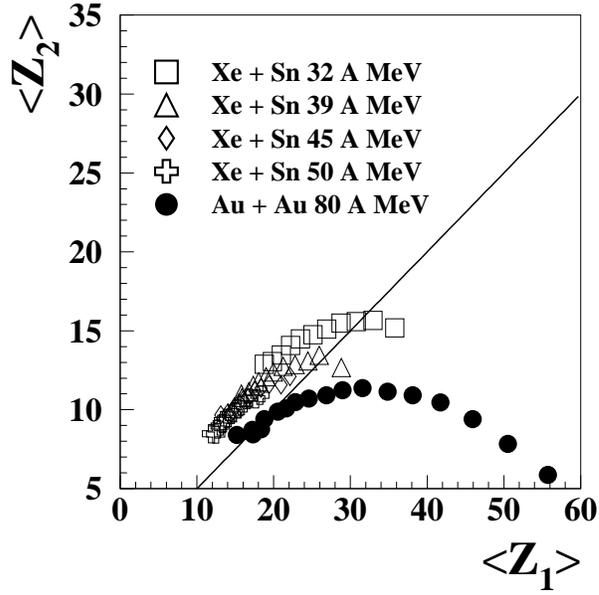}
\caption{\it Average size of the second largest fragment 
$<$Z$_{2}$ $>$ 
as a function of the average size of the largest fragment $<$Z$_{1}$ $>$  
for Xe+Sn at  32, 39, 45  and 50~A~MeV mono-sources and for 
quasi-projectiles measured in Au+Au reactions at 80~A~MeV. The line corresponds 
to $<$Z$_{2}$ $>$ $=$ $<$Z$_{1}$ $>$ $/$2.}
\label{zbig12} 
\end{center}
\end{figure}

\indent
Besides this global analysis it is interesting to look at which energy 
$<$Z$_{2}$$>$ is exactly equal to $<$Z$_{1}$$>$$/$2. Indeed it has been shown in 
\cite{Ma05} that for the Ar quasi-projectile, the relationship between 
$<$Z$_{1}$$>$ and $<$Z$_{2}$$>$ has a bell shape. 
The maximum is at $<$Z$_{2}$$>$ $=$ $<$Z$_{1}$$>$$/$2 
and is observed at an excitation energy E$^{*}$ for which the Zipf law 
is verified.
Figure \ref{zbig12} shows the relationship between the two largest fragments 
measured in peripheral Au QP (filled circles) and in Xe+Sn mono-sources
(open symbols). The line represents the locus where 
$<$Z$_{2}$$>$$=$$<$Z$_{1}$$>$$/$2.
A bell shape is observed for Au QP events. 
At the maximum $<$Z$_{1}$$>$$\simeq$30
and $<$Z$_{2}$$>$$\simeq$12: These values refer to an excited source at 
E$^{*}$=6~A~MeV 
(see figure \ref{zbigEstar} and figure \ref{zbig2Estar})
significantly below E$_{Zipf}$=8.5~A~MeV. However above 
E$^{*}$=8~A~MeV, the relation 
$<$Z$_{2}$$>$ $=$ $<$Z$_{1}$$>$$/$2 is verified. Thus our data on Au 
quasi-projectiles do not lead to the coherent picture seen in the 
fragmentation of an Ar quasi-projectile \cite{Ma05}.
For Xe+Sn mono-sources at 32~A~MeV, $<$Z$_{2}$$>$$=$$<$Z$_{1}$$>$$/$2 is verified 
only at the maximum but it corresponds to E$^{*}$= 5~A~MeV, a value far below 
E$_{Zipf}$=7.5~A~MeV for mono-sources. In the other Xe+Sn mono-sources, 
we do not find data supporting the coherence seen in \cite{Ma05}. 
At this stage of the analysis, one could suggest a possible influence of the 
size of the sources between Ar quasi-projectiles and the present systems 
which are more than four times larger.

\subsection{Fragment size fluctuations}

\begin{figure}[!hbt]
\begin{center}
\includegraphics*[scale=0.45]{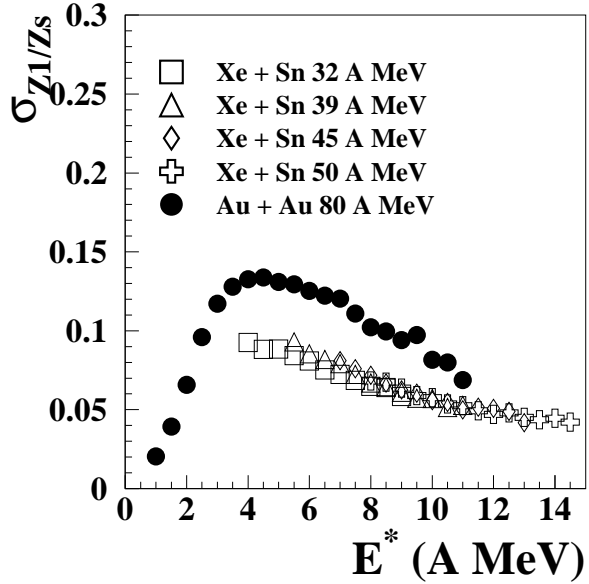}
\caption{\it Fluctuations of the average size of 
the largest fragment normalized to the source size Z$_{1}$/$Z_{s}$
of the multifragmenting system
as a function of the excitation energy for Xe+Sn mono-sources 
and Au QP events.}
\label{fluctuations} 
\end{center}
\end{figure}

\indent
It is expected that close to the critical point fluctuations 
in the cluster size distribution are maximum. 
It is suggested in various models that the 
fluctuations in the size of the largest fragment could be a good indicator of 
the distance of the system from the critical point, even for finite systems. 
In the experimental study of ref.
\cite{Ma05} at the excitation energy where  Zipf law is verified the authors 
observe also the largest value of the normalized fluctuations . 
In figure \ref{fluctuations} are reported these fluctuations, 
$\sigma_{Z_{1}/Z_{s}}$=$\sqrt{<(Z_{1}/Z_{s}){^2}>-<(Z_{1}/Z_{s})>{^2}}$,
as a funtion of E$^{*}$ as deduced from 
the present data set. 
For Xe+Sn mono-sources $\sigma_{Z_{1}/Z_{s}}$ continuously decrease. 
For Au QP $\sigma_{Z_{1}/Z_{s}}$ reaches 
a maximum value around $\simeq$4.5~A~MeV. The highest value of 
$\sigma_{Z_{1}/Z_{s}}$, $\simeq$0.135, is in good 
agreement with other data \cite{elliott2} (see also table 3 in "Fluctuations 
of fragment observables" in \cite{WCI}). 
Indeed in quasi-projectiles measured at 35~A~MeV in Au+Au collisions
with Multics, a $\sigma_{Z_{1}/Z_{s}}$ value of 0.14 is obtained and 
$\sigma_{Z_{1}/Z_{s}}$$\simeq$0.12-0.13 for EOS data. 
The maximum is observed at an excitation energy very close to 
E$_{crit}$ extracted from the Fisher procedure and well below 
E$_{Zipf}$ $\simeq$8.5~A~MeV extracted from the Zipf law fit. 
A striking feature is the superposition of all Xe+Sn data whatever the 
bombarding energy which confirm that the fragmentation pattern 
in Xe+Sn central collisions depends only on the
excitation energy. Finally, the normalized fluctuations are systematically 
smaller for Xe+Sn mono-sources than for Au QP. 
It is very interesting to remark that, for a given value of deposited
excitation energy, the Au QP data systematically show a lower degree of
fragmentation. This is measured by the lower fragment multiplicity,
figure \ref{Multiplicitydist},
the higher charge asymmetry between the two heaviest fragments, figures 
\ref{zbigEstar} and \ref{zbig2Estar}
and the higher fluctuation, figure \ref{fluctuations}, associated to the 
Au QP data. This experimental finding points towards a different fragmentation
scenario between the two data sets characterized by different entrance
channels. Different physical reasons may be invoked: a different
thermodynamic path in the temperature-pressure plane, an influence of the
radial flow dynamics on the fragmentation pattern... Further studies will be 
needed to disentangle between the different interpretations
and to investigate the influence of the asymmetry in the entrance channel and 
to explore a different size region.  

\subsection{Configurational energy and heat capacity}

\begin{figure}[!hbt]
\begin{center}
\includegraphics*[scale=0.45]{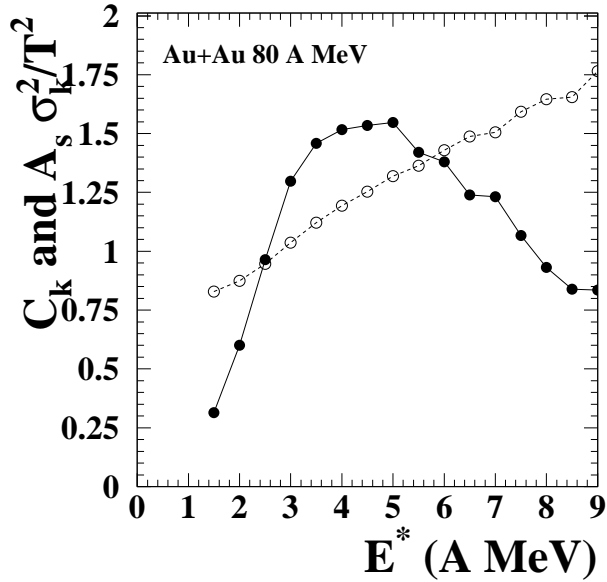}
\caption{\it  Normalized fluctuations $A_{s}\sigma_{k}^{2}/T^{2}$ 
(filled circles) and kinetic energy capacity  $c_{k}$  
(open circles) as a function of excitation energy for Au 
QP reactions at 80~A~MeV.}    
 \label{cneg_auau} 
\end{center}
\end{figure}

\indent  
For a finite system in the liquid-gas coexistence zone one
expects large fluctuations of the total kinetic energy 
(supposed only thermally coupled to the system)
leading to a negative
value of the heat capacity \cite{chomaz1,chomaz2,FrancescaAnal,Gross}.
For equilibrated excited nuclear systems one assumes
that the total excitation energy E$^{*}$ can be separated in two components, 
$E^*=E_{k}+E_{pot}$, where
E$_{k}$ and E$_{pot}$ are the total kinetic energy and total configurational 
energy respectively. The total heat capacity is defined as 
$c_{tot}=c^{2}_{k}/(c_{k}-A_{s}\sigma_{k}^{2}/T^{2})$ where $c_{k}=dE_{k}/dT$; 
the temperature T is estimated by solving the kinetic equation of state
\cite{nicolasthese,michela4} 
and $A_{s}$ is the size of the source. In multifragmentation studies the 
total kinetic energy E$_{k}$ should be determined at the freeze-out stage but 
such configuration  is not experimentally accessible. However, one may
deduce the heat capacity from the configurational energy
of measured partitions, if side feeding effects are properly 
accounted for. The general procedure has been thoroughly checked and well 
developed in \cite{michreliability}.\\ 

\begin{figure}[!hbt]
\begin{center}
\includegraphics*[scale=0.45]{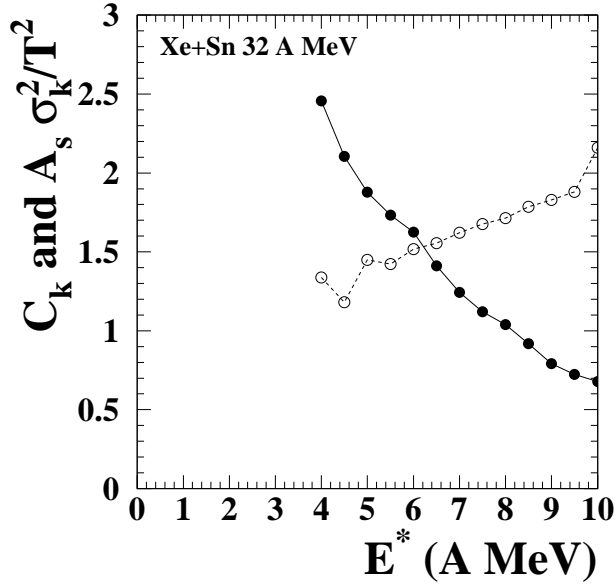}
\caption{\it  Normalized fluctuations $A_{s}\sigma_{k}^{2}/T^{2}$ 
(filled circles) and kinetic energy capacity  $c_{k}$  
(open circles) as a function of excitation energy for Xe+Sn mono-source events
at 32~A~MeV.}    
 \label{cneg_xesn} 
\end{center}
\end{figure}

\indent  
Results of the analysis are presented in figure \ref{cneg_auau} 
for Au QP. Normalized fluctuations $A_{s}\sigma_{k}^{2}/T^{2}$ 
(filled circles in figure \ref{cneg_auau}) show a bell 
shape as a function of the excitation energy  E$^{*}$ with a maximum around
$\simeq$4~A~MeV. It is worth noticing that for Au QP events, 
fluctuation of the average size
of the largest fragment, normalized to the source size ${Z_{1}/Z_{s}}$, is 
maximum at roughly the same energy.
$c_{k}$ (open circles in figure \ref{cneg_auau}) monotonically increases 
with the excitation energy and crosses the normalized fluctuations at 
E$^{*}$$\simeq$2.5~A~MeV and E$^{*}$$\simeq$5.5~A~MeV. 
In the hypothesis of thermal equilibrium such crossings are associated with two 
divergences  and a negative branch in the total heat capacity \cite{chomaz1}. 
It has been shown in the literature, (\cite{WCI} p.259 figure 8)
and \cite{pic06}, that the magnitude of the fluctuations decreases
when dynamical effects are present. As we have shown in figure 
\ref{distrithetaflot}, the sample selected with a compacity criterion 
exhibits a flat cos$\theta_{flow}$ distribution which indicates that a strong 
degree of equilibrium is reached. Results shown in figure \ref{cneg_auau}
are compatible with those found in Au QP for the same data at
another selection of compact events  \cite{pic06} and 
by the MULTICS collaboration \cite{michela2,michreliability,michela4}
but at 35~A~MeV. Moreover, locations 
of divergences and maximal normalized fluctuations are observed at the same 
excitation energies in both INDRA and MULTICS data. Figure \ref{cneg_xesn}  
shows the  result of the analysis for Xe+Sn mono-sources at 32~A~MeV. 
Due to the limited range of E$^{*}$, only the divergence at high excitation 
energy is observed but it is worth noticing that it is roughly at the same 
E$^{*}$ as in the Au QP events. As the incident energy increases 
the position of the second divergence remains constant but the negative part of the
total heat capacity progressively disappears as we are exploring
higher excitation energies.
 
\section{ Discussion}

\subsection{Summary of experimental findings on scaling, ordering and 
fluctuations of fragment observables.}
Let us first recall our findings regarding pseudo-critical behaviours and 
fluctuations of fragment observables in Xe+Sn mono-sources and
Au QP events. Applying the Fisher procedure a good 
scaling of the cluster yields is observed, 
providing the pseudo-critical energy E$_{crit}$ 
where a power law is observed, the value of the critical
exponents $\tau$, in the range 2.09-2.56 and $\sigma$$\simeq$2/3. 
For a liquid-gas transition $\tau$=2.21 for Ising 3D 
\cite{hasenbusch}, for a mean field Lattice Gas
$\tau$=2+1/D (=2.33 for a 3 dimensional Lattice). In an infinite 3D
percolation $\tau$=2.18 \cite{stauffer}
and $\tau$=2.2$\pm$0.1 in a finite one \cite{Aldo}.
The value of $\tau$ does not provide a discriminating test for 
characterizing the universality class of the transition. 
In the present work the values of the critical energy are in the range 
3.8-4.5~A~MeV. Conversely, a Zipf law is observed for all systems 
but at an energy higher than E$_{crit}$:
E$_{Zipf}$$\sim$7.5~A~MeV for Xe+Sn mono-sources and E$_{Zipf}$$\sim$8.5~A~MeV 
for Au QP events. The normalized fluctuations of the largest 
fragment exhibit a maximum around 4.5~A~MeV. Finally, for the 
Au QP, the normalized kinetic fluctuations 
$A_{s}\sigma_{k}^{2}/T^{2}$ reach a maximum around 4~A~MeV. 
Within the framework of thermodynamic equilibrium, the heat capacity is negative
between E$^{*}$=2.5 and E$^{*}$=5.5~A~MeV.
The present results agree with many works reporting 
numerous signatures compatible with a system in a mixed phase of liquid and 
vapor
(\cite{michela2,remi00,nicolasthese,michreliability,I31-Bor01,I40-Tab03,pocho,borderiephysg,pic06,Tam05,Bonnet05,RIVET98,I29-Fra01,I57-Tab05,I51-Fra05}).\\

\indent
Several conclusions can be drawn out from our findings. First, for Au QP 
and Xe+Sn mono-sources, the "critical energies" extracted 
from Fisher and Zipf procedures are not compatible with 
each other. Such a conclusion is at variance with ref.\cite{Ma05}. 
Moreover, in our data, the Zipf law is 
observed at an excitation energy well above that where the
fluctuations connected to the heat capacity are maximum. 
It was recently stated \cite{campi05} 
that the Zipf law could be deduced from the power law behaviour of the 
cluster size distribution and the 
exponents $\tau$ and $\lambda$ would roughly satisfy
$\lambda$=1/($\tau$-1) at the critical point, leading to $\tau$=2, a value 
below our experimental values and from most of the theoretical calculations 
(for example 3D percolation or Lattice Gas Model). In our data, 
$\tau$=2.2 gives $\lambda$=0.86, a value obtained at  E$^{*}$=9.5~A~MeV, higher 
than the excitation energy where Zipf law is verified and consequently 
even further from the critical energy obtained from the Fisher fitting 
procedure.\\ 

\indent
A more interesting remark is that the critical energy extracted from the
Fisher procedure is located in the domain where the normalized
fluctuations of the largest fragment and the configurational energy fluctuations 
reach their highest values.
This was clearly deduced from the Au QP, and could be 
only extrapolated from  Xe+Sn mono-sources data since in such cases low 
excitation energies are not explored.\\

\indent
The Zipf law is not a reliable way to identify a system close to or at the 
critical point. Further investigations are needed to explain differences 
between our results and those shown in \cite{Ma05}. Size effects might be a 
possible direction to explore. Another interesting proposition from Bauer and 
collaborators is to replace the Zipf law by the Zipf-Mandelbrot law 
\cite{bauer06}.\\

\indent
The results of Fisher scaling and negative heat capacity do not show any 
dependence on the type of collisions studied. Conversely, some differences 
appear in the energies where the Zipf law is verified and in the fluctuations of 
the largest fragment. This observation points to the specific role of the largest
fragment as already observed in \cite{I57-Tab05,I51-Fra05}, which appears more
sensitive to the fragmentation mechanism.
Indeed these results suggest that for Au QP and Xe+Sn mono-sources
the fragmentation mechanisms are different.\\
 
\indent
To go further in the discussion we used predictions of theoretical models as 
guidelines to understand our results.

\subsection{Guidelines from the Lattice Gas Model}
This model is well known to describe a first as well as a second order phase 
transition. Since thermodynamical
conditions of the model are well defined (temperature, energy, density, 
coexistence line, critical point) it is straightforward to verify whether a 
Fisher scaling is observed in the Lattice Gas Model, and how to interpret 
the "Fisher critical point" in such a framework. A Fisher procedure 
\cite{francescalat} was performed at different densities (critical, 
subcritical and supercritical).
A very good scaling is observed for all densities but the critical temperature 
extracted from the fit is density dependent. 
The analysis of ref. \cite{francescalat} is analogous to the one we have 
carried out with the data. In this case, it would be better to describe the derived 
power law as a pseudo critical behaviour since it is not only observed at 
the critical point but also inside as well as outside the coexistence region.
Gathering the results in a temperature-density plot, one finds that the 
locus of the pseudo-critical points is close to the so called Kertesz's line 
\cite{Kertesz,campi99}. 
Thus the observation of a Fisher scaling is not a tool to determine 
the location of the critical point. 
However, the Fisher scaling is compatible with fragmentation 
inside the coexistence region, in agreement with our experimental 
indications from configurational energy fluctuations.
Moreover the narrow range of E$^{*}_{crit}$ in our analysis suggests, in the
framework of the Lattice Gas Model, that a narrow range of density is explored
by the studied systems.

\subsection{Guidelines from the Statistical Multifragmentation Model SMM}

\begin{figure}[!hbt]
\begin{center}
\includegraphics*[scale=0.45]{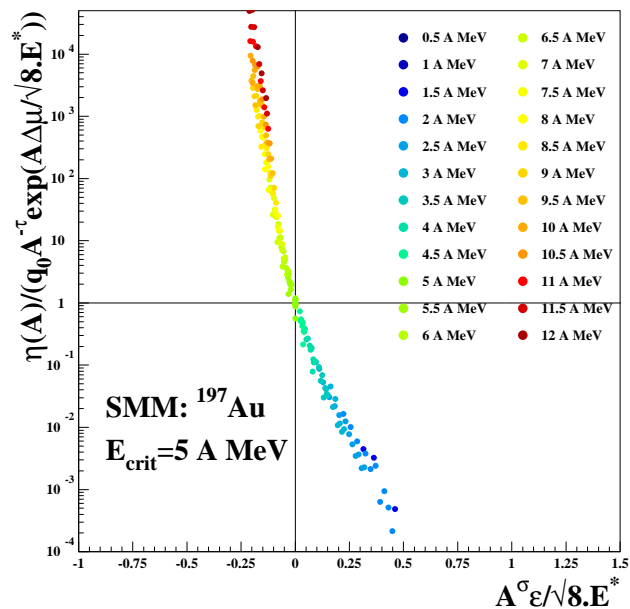}
\caption{\it Fisher scaling for SMM events.}
\label{fishersmm} 
\end{center}
\end{figure}

We have performed a Fisher procedure on the predictions of a well known 
statistical multifragmentation model (SMM). A detailed description of SMM can 
be found in \cite{bondorf}. The basic assumption of the model is
the statistical distribution of the breakup channels of a decaying system at 
thermal equilibrium. SMM was largely used to reproduce the characteristics of 
multifragmentation
events \cite{remi00,michela1}. In the present work, calculations have 
been performed, for a fixed value of the source (Z$_{0}$=79,
A$_{0}$=197) at two freeze-out volumes ($V_{fo}$= 3$V_{0}$ and 6$V_{0}$) and 
for a flat distribution of excitation energy in the range
0.5$\leq$E$^{*}$$\leq$12~A~MeV. The experimental filter has not been 
applied to the predictions of the model since our goal was not to reproduce the
data.\\

\begin{figure}[!hbt]
\begin{center}
\includegraphics*[scale=0.45]{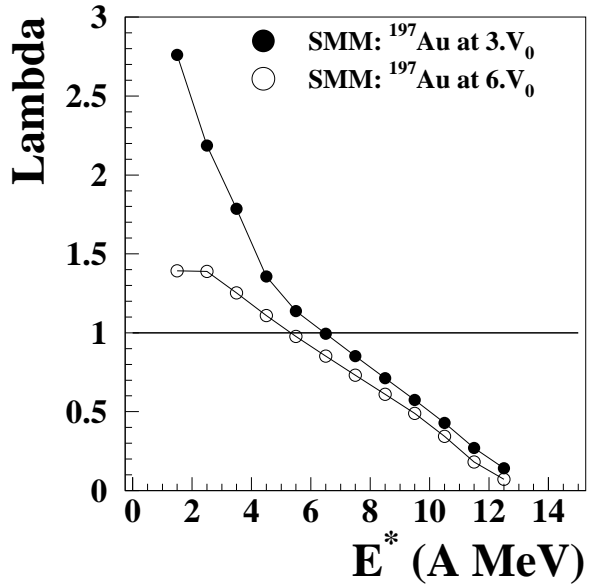}
\caption{\it Zipf analysis for SMM calculations.}
\label{zipfsmm} 
\end{center}
\end{figure}

\indent
We first applied the Fisher procedure to the calculations. The yield 
parameterization is the same as the one used to study the experimental data 
(see section 3).
Predictions at $V_{fo}$= 3$V_{0}$ are shown in figure \ref{fishersmm} where
$\eta_{A}/(q_{0}A^{-\tau})\times exp(A\Delta\mu/\sqrt{8.E^{*}})$ is plotted against 
$\varepsilon A^{\sigma}/\sqrt{8.E^{*}}$. 
Here for the caloric curve we assume the same ansatz as for the 
experimental data. In the framework of SMM we have access to an average 
temperature corresponding to a given excitation energy. 
We have observed that in our calculations the caloric curve 
(similar to the one shown in figure 1 in \cite{bondorf2})
is close to the one obtained in \cite{michela724}.
It was demonstrated in \cite{michela724} that the values of the extracted 
critical parameters do not depend on the shape of the caloric curves.
Results of the calculations 
fall on a single curve, which is not a straight line at variance with the
experimental data (indeed in SMM the c$_{0}$ coefficient deduced from the fit
procedure evolves rapidly with E$^{*}$, whereas it appears roughly constant in
the data) and thus a scaling law is observed for SMM events. 
Fisher "critical point" is obtained at E$^{*}\simeq$5~A~MeV.
This model presents a phase transition of first order \cite{Gross}, meaning 
that in this case the power law behaviour as well as the scaling 
cannot be connected to the location of a critical point. They rather signal 
a maximal fluctuation point. Indeed under the microcanonical constraint 
kinetic energy fluctuations are maximum when the configuration energy has the 
largest spread, i.e. when the size distribution is close to a power law. 
Therefore, in the statistical multifragmentation 
model, the Fisher "critical point" can be interpreted as the energy where 
fluctuations are maximum. This is in agreement with our experimental 
findings.
Finally, the analysis performed at $V_{fo}$= 6$V_{0}$ does not indicate a 
noticeable freeze-out volume influence on the scaling properties nor on the 
parameters extracted from the Fisher fit ($E_{crit}$, $\tau$ and 
$\sigma$).\\

\begin{figure}[!hbt]
\begin{center}
\includegraphics*[scale=0.45]{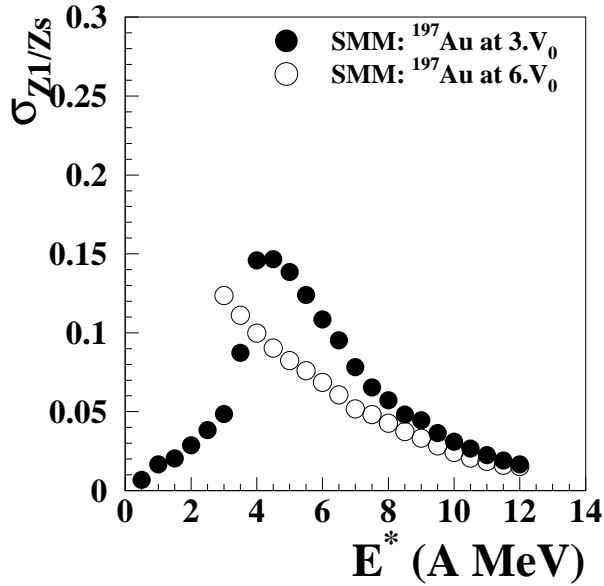}
\caption{\it Fluctuations of the average size of 
the largest fragment normalized to the source size Z$_{1}$/$Z_{s}$
as a function of the excitation energy for SMM calculations.}
\label{fluctnormzszbib} 
\end{center}
\end{figure}

\indent
The Zipf law fit has also been performed on SMM events using the 
same protocol as for experimental data. Results of the analysis are 
presented in figure \ref{zipfsmm} 
for $V_{fo}$= 3$V_{0}$ (filled circles) and 6$V_{0}$ (open circles). 
For both freeze-out volumes we observe a decrease of the 
$\lambda$ parameter as the excitation energy E$^{*}$ increases. 
When $V_{fo}$= 3$V_{0}$,  
$\lambda$=1 is reached at E$^{*}$=6.5~A~MeV, a value which falls in the region 
close to the second divergence of the heat capacity. At 
variance with the Fisher fitting procedure, the Zipf law fit seems to evolve 
with the freeze-out volume. As indicated in figure \ref{zipfsmm}, one observes a 
shift towards lower excitation energy as $V_{fo}$ increases. 
The value found in this case, E$_{Zipf}$=5~A~MeV, is comparable with the one 
extracted from the Fisher procedure, E$_{crit}$$\simeq$4.5~A~MeV. 
All the trends and values shown in figure \ref{zipfsmm} are similar when the 
analysis is performed for fragments at the freeze-out stage. 
We can mention that the same volume effect was seen in \cite{Ma99} for the
occurrence of the Zipf law as a function of the density in the Lattice Gas
Model. Values of E$_{Zipf}$ obtained from SMM fall below the experimental ones 
for both sets of data. 
Figure \ref{zipfsmm} however suggests that the present Au QP data may be 
associated with smaller freeze-out volumes. To investigate the potential role of 
the freeze-out volume on the Zipf procedure, one
needs other theoretical tools.\\ 

\indent
Figure \ref{fluctnormzszbib} represents the normalized fluctuations of the 
charge of the largest fragment $\sigma_{Z_{1}/Z_{s}}$ 
as a funtion of E$^{*}$ deduced from SMM calculations for the two volumes. 
Because of the presence of fission, the points for $V_{fo}$=6$V_{0}$ are not
displayed below 3~A~MeV in figure \ref{fluctnormzszbib}.
The observed trend is very similar to the one seen in figure 
\ref{fluctuations} which again indicates a smaller volume for Au QP.
It has been shown indeed that the 
volume influences the size of the biggest fragment \cite{des98}.
More specifically, the normalized fluctuations reach a maximum value at 
E$^{*}$=4.5~A~MeV for a source at $V_{fo}$=3$V_{0}$ and the magnitude is 
around 0.15, values very close to what has been deduced in experimental 
data for Au QP. 

\section{Conclusions}

To conclude, a study of the features of multifragmenting sources formed in 
Xe+Sn mono-sources between 32 and 50~A~MeV and Au QP 
at 80~A~MeV has been reported. Scaling law of the cluster yields, 
ranking in the average size of fragment and fluctuations in cluster sizes 
and kinetic observables have been used to characterize the fragmentation 
process over a wide range of excitation energy.\\

\indent
From the analysis of cluster yields, we have shown that all data follow a 
Fisher scaling which coherently points to an excitation energy of 
$\simeq$4~A~MeV associated with a power law distribution. This "critical point" corresponds 
closely to the region of maximum fluctuations both of the configurational energy
and the charge of the largest fragment. 
Theoretical models suggest that such point may not be associated with a critical 
behaviour, but also be characteristic of a subcritical behaviour (coexistence 
region of a first order phase transition). The high value of configurational 
energy fluctuations observed in our data set tends to confirm this 
interpretation.
Data discussed in this work are compatible with a subcritical phenomenon. 
This interpretation of our results also
agrees with previous analyses, on the same systems, 
about other signals of phase transition like spinodal decomposition and 
bimodality \cite{I40-Tab03,pic06,Tam05,Bonnet05}.\\ 

\indent
A Zipf-like analysis on the average cluster charge has been carried out and 
indicates that the Zipf law is observed at an excitation
energy which does not coincide with the critical energy deduced from the other
analyses. A coherent picture  from various criticality 
signals and Zipf law is not observed neither for Xe+Sn mono-sources 
nor Au QP.


\small{

} 

\end{document}